\documentclass[12pt]{article}

\begin{document}
\begin{titlepage}

\title{Regularized expression for the gravitational 
energy-momentum in teleparallel gravity and the principle
of equivalence}

\author{J. W. Maluf$\,^{*}$, M. V. O. Veiga\\
Instituto de F\'{\i}sica, \\
Universidade de Bras\'{\i}lia\\
C. P. 04385 \\
70.919-970 Bras\'{\i}lia DF, Brazil\\
and\\
J. F. da Rocha-Neto\\
Universidade Federal do Tocantins\\
Campus Universit\'ario de Arraias\\
77.330-000 Arraias TO, Brazil\\}
\date{}
\maketitle

\begin{abstract}
The expression of the gravitational energy-momentum defined in the
context of the teleparallel equivalent of general relativity 
is extended to an arbitrary set of real-valued tetrad fields, by
adding a suitable reference space subtraction term. 
The characterization of tetrad fields as reference frames is 
addressed in the context of the Kerr space-time. It is also pointed
out that Einstein's version of the principle of equivalence does 
not preclude the existence of a definition for the gravitational 
energy-momentum density. 
\end{abstract}
\par
\bigskip

\thispagestyle{empty}
\par
\bigskip
\noindent (*) e-mail: wadih@fis.unb.br\par
\end{titlepage}
\newpage

\noindent
\section{Introduction}
The notion of gravitational energy-momentum has been adressed 
recently in the framework of the teleparallel equivalent of general 
relativity (TEGR) 
\cite{Maluf2,Maluf3,Maluf4,Maluf5}. 
The TEGR is an alternative geometrical 
description of Einstein's general relativity 
in terms of tetrad fields $e^a\,_\mu$ and of the torsion tensor
$T^a\,_{\mu\nu}=\partial_\mu e^a\,_\nu - \partial_\nu e^a\,_\mu$
($a,b,,\,...$ and $\mu,\nu,\,...$ are SO(3,1) and space-time indices,
respectively). The field equations for the tetrad field $e^a\,_\mu$
are precisely equivalent to Einstein's equations. Therefore it is not
a new theory for the gravitational field. The torsion tensor is 
related to the antisymmetric part of the Weitzenb\"ock connection
$\Gamma^\lambda_{\mu\nu}=e^{a\lambda}\partial_\mu e_{a\nu}$. The 
curvature tensor constructed out of this connection vanishes 
identically. Therefore this connection allows the notion of
distant parallelism in space-time. Of course one may 
construct the Christoffel symbols and consider the physical
and geometrical properties of the (nonvanishing) Riemann-Christoffel 
tensor.

The geometrical framework determined by the tetrad field and 
torsion tensor has proven to be suitable to investigate the problem 
of defining the gravitational energy-momentum.
A consistent expression developed in the realm of the TEGR shares
many features with the expected definition. The 
gravitational energy-momentum $P^a$ \cite{Maluf2,Maluf3} obtained in 
the framework of the TEGR has been investigated in the
context of several disctinct
configurations of the gravitational field. For asymptotically flat 
space-times $P^{(0)}$ yields the ADM energy \cite{ADM}. In the context 
of tetrad theories of gravity, asymptotically flat space-times may be
characterized by the asymptotic boundary condition,

\begin{equation}
e_{a\mu}\cong\eta_{a\mu}+{1\over 2}h_{a\mu}({1/r})\,,
\label{1}
\end{equation}
and by the condition $\partial_\mu e^a\,_\nu =O({1/r^2})$ in the 
asymptotic limit $r\rightarrow \infty$.  
In the asymptotic limit above the quantity $\eta_{a\mu}$ in Eq. (1)
coincides with the metric tensor of the Minkowski space-time 
$\eta_{ab}=(-+++)$. An important property of tetrad fields that 
satisfy the condition above is that in the flat space-time limit we
have $e^a\,_\mu(t,x,y,z)=\delta^a_\mu$, and therefore 
$T^a\,_{\mu\nu}=0$. Hence for the flat space-time we normally 
consider a set of tetrad fields such that $T^a\,_{\mu\nu}=0$
in any coordinate system. This condition establishes the
reference space.  However, 
in general an arbitrary set of tetrad fields that yields the metric
tensor for asymptotically flat space-times does not satisfy the
asymptotic condition given by Eq. (1). Moreover for such tetrad
fields we have in general $T^a\,_{\mu\nu}\ne 0$ in the flat 
space-time. It might be argued, therefore,
that the expression for the gravitational energy-momentum mentioned
above is restricted to a particular class of tetrad fields, namely,
to the class of frames such that $T^a\,_{\mu\nu}=0$ if $e^a\,_{\mu}$
represents the flat space-time tetrad field.

The definition $P^a$ is invariant under global SO(3,1) transformations.
We have argued elsewhere \cite{Maluf3,Maluf4,Maluf5} that it makes 
sense to have a dependence of $P^a$ on the frame. The energy-momentum
in classical theories of particles and fields does depend on the frame,
and we assert that such dependence is a natural property of the 
gravitational energy-momentum. The total energy of a relativistic
body, for instance, depends on the frame.
We normally assume that a set of tetrad fields is adapted to an
ideal observer in the space-time determined by the metric tensor 
$g_{\mu\nu}$. For a given gravitational field configuration (a
black hole, for instance), the infinity of possible observers is
related to the infinity of tetrad fields (related by a local SO(3,1)
transformation) that yields the metric tensor $g_{\mu\nu}$. 
Let $x^\mu(s)$ denote the worldline $C$ of an observer,
and $u^\mu(s)=dx^\mu/ds$ its velocity along $C$. We may identify
the observer's velocity with the $a=(0)$ component of $e_a\,^\mu$,
where $e_a\,^\mu e^a\,_\nu=\delta^\mu_\nu$. Thus, 
$u^\mu(s)=e_{(0)}\,^\mu$ along $C$ \cite{Hehl}. 
The acceleration of the observer is given by 

\begin{equation}
a^\mu= {{Du^\mu}\over{ds}}={{De_{(0)}\,^\mu }\over{ds}}=
u^\alpha \nabla_\alpha e_{(0)}\,^\mu\,.
\label{2}
\end{equation}
The covariant derivative
is constructed out of the Christoffel symbols. We see that $e_a\,^\mu$
determines the velocity and acceleration along the worldline of an 
observer adapted to the frame.
From this perspective we conclude that a given set of 
tetrad fields, for which $e_{(0)}\,^\mu$ describes a congruence of
timelike curves, is adapted to a particular class of observers,
namely, to observers determined by the velocity field 
$u^\mu=e_{(0)}\,^\mu$, endowed with acceleration $a^\mu$. If 
$e^a\,_\mu \rightarrow \delta^a_\mu$ in the limit 
$r \rightarrow \infty$, then $e^a\,_\mu$ is adapted to stationary
observers at spacelike infinity. We may say, therefore, that $P^{(0)}$
yields the ADM energy for such observers.

In this article we will extend the definition $P^a$ for the 
gravitational energy-momentum previously considered for arbitrary 
tetrad fields, namely, for tetrad fields that satisfy 
$T^a\,_{\mu\nu}\ne 0$ in the flat space-time. The redefinition is
the only possible consistent extension of $P^a$, valid for tetrad 
fields that do not satisfy boundary conditions like Eq. (1). We will 
also argue that an existing version of the principle of equivalence, 
namely, Einstein's version of the principle, does not pose any 
obstacle to the concept of localized gravitational energy. We will 
show that the usual (textbook) version of the principle was never 
accepted by Einstein.

\section{The principle of equivalence and the localizability of the
gravitational energy}

The concept of energy in classical electrodynamics is very 
simple and well known. We consider an arbitrary volume in the 
three-dimensional space and verify the existence of field lines of 
the electric and/or magnetic field in this region. The 
electromagnetic energy density is given by the 
standard expression that consists of the sum of the square of the 
electric and magnetic fields, and therefore is nonvanishing in a
space-time region where the field lines are present.
Charged particles in this region 
experience the Lorentz force. Therefore the manifestation of the
Lorentz force is an indication of the existence of electromagnetic 
energy density.

Unfortunately there is not a simple picture in general relativity that
relates gravitational ``field lines" to the existence of gravitational
energy density. Nevertheless, it is legitimate to expect that 
the manifestation of gravitational forces in a three-dimensional region 
is an indication of the existence of gravitational energy-momentum 
density in this region. However in general relativity there is a point 
of view according to which the gravitational energy density cannot be 
localized (see \cite{MTW} \S 20.4). The argument is the following.
In any given small region of the space-time manifold we can find a 
coordinate system such that the Christoffel symbols disappear. In terms 
of this appropriate coordinate system the small region in question is
``free of gravitational fields". In summary, this is the argument that 
has been endorsed by many authors, who claim that the nonlocalizability 
of the gravitational energy-momentum is due to the principle of 
equivalence.

We do not endorse the conclusion above for various reasons. First, it is
well known that the vanishing of the Christoffel symbols does not imply 
the vanishing of tidal forces in any infinitesimal region of the 
space-time. Therefore, the assertion that this region is free of 
gravitational fields is questionable, because we should agree that the
existence of gravitational forces (e.g., on the worldline of a particle)
is due to gravitational fields. It is not reasonable to accept the idea 
of having a force in a space-time region without the associated field
in this same region.

Second, the principle of equivalence that supports the conclusion above is
related to Pauli's version of the principle \cite{Pauli},
but is different from Einstein's version. According to 
Pauli's formulation, {\it for every infinitely small world region
there always exist a coordinate system in which gravitation has no
influence either on the motion of particles or any other physical 
processes}. The distinction between Einstein's and Pauli's formulation 
of the principle of equivalence has been addressed by Norton
\cite{Norton}. From the point of view of Pauli's formulation, the
vanishing of the Christoffel symbols in a space-time region implies that 
gravitation has no effect in this region. We know, however, that what 
really vanishes in such region are the first derivatives of the metric 
tensor. In our opinion the mathematical feature that 
consists of the vanishing of the first derivatives of {\it any} metric 
tensor - but not of the second and highest derivatives - along any 
worldline in a Riemannian or pseudo-riemannian manifold, in any 
dimension, cannot be taken as a physical principle. It is just a feature
of differential geometry.

Paulis's formulation of the principle of equivalence is different 
from Eintein's formulation \cite{Norton}. The latter is
unquestionably considered to be the breakthrough that 
led Einstein to establish the conditions under which a noninertial frame 
is equivalent to an inertial one, extending in this way the principle of
relativity. In view of the practical difficulties related to the 
description of arbitrary gravitational fields by means of the principle 
of equivalence, the latter was abandoned in favour of the principle of
general covariance. Nevertheless 
the importance of the principle has always been recognized by Einstein 
in the years after the formulation of general relativity \cite{Norton}.

Einstein's version of the principle of equivalence \cite{Norton} consists 
of considering a reference frame $K$ (a ``Galilean system"), and a 
reference frame $K^\prime$, which is uniformly accelerated with respect 
to $K$. Then one asks whether an observer in $K^\prime$ must understand
his condition as accelerated, or whether there remains a point of view
acording to which he can interpret his condition as at ``rest". Einstein
concludes that by assuming the existence of a homogeneous gravitational 
field in $K^\prime$ it is possible to consider the latter as at rest.
In his words: {\it The assumption that one may treat $K^\prime$ as at
rest, in all strictness without any laws of nature not being fulfilled
with respect to $K^\prime$, I call the `principle of equivalence'} 
\cite{Norton}.

Again considering Ref. \cite{Norton}, we observe that an important
remark about Einstein's formulation of the principle of equivalence is 
not widely considered in the literature: Einstein's formulation is 
established in Minkowski space-time. The passage from $K$ to $K^\prime$
amounts to a frame transformation in a finite region of the space-time,
not a coordinate transformation. Moreover Einstein never endorsed 
Pauli's formulation. Einstein objected that {\it in the infinitely small 
every continuous line is a straight line} \cite{Norton}. He believed 
that the restriction to infinitesimal regions makes it impossible 
to distinguish the geodesic world lines of free point masses from other
world lines and thus it is impossible to judge whether - in the words of
Pauli's formulation - ``gravitation has no influence on the motion of 
particles". Quoting Norton \cite{Norton}: {\it It has rarely been 
acknowledged that Einstein never endorsed the principle of equivalence
which results, here called the ``infinitesimal principle of equivalence".
Moreover, his early correspondence contains a devastating objection to
this principle: in infinitesimal regions of the space-time manifold it is
impossible to distinguish geodesics from many other curves and therefore
impossible to decide whether a point mass is in free fall}.
\footnote{
It is worthwhile to point out a compact statement of the principle 
formulated by Einstein in 1918 \cite{Norton}: 
{\it Principle of Equivalence: inertia and gravity
are wesensgleich (identical in essence). From this and from the results
of the special theory of relativity it necessarily follows that the 
symmetrical tensor $g_{\mu\nu}$ determines the metrical properties of 
space, the inertial behaviour of bodies in it, as well as the 
gravitational action}.}

The principle of equivalence follows from the equality of inertial and
gravitational masses, and really establishes the equivalence between a 
noninertial reference frame and an inertial one with the addition of a 
suitable gravitational field. The nonvanishing of tidal forces in 
infinitesimal regions of the space-time does not allow us to conclude 
that such regions can be made free of gravitational fields by means of
coordinate transformations. Two nearby particles in free
fall undergo geodesic deviation irrespective of whether the metric 
tensor is reduced to the Minkowski form along their worldlines. 
We recall that by means of coordinate 
transformations one cannot reduce the tetrad field or the torsion tensor 
at a space-time point to their flat space-time values.
Any space-time region, infinitesimal or not, is flat and consequently
``free of gravitational fields" if and only if the Riemann-Christoffel
tensor vanishes in this region.
Arguments based on the ``infinitesimal
principle of equivalence" are not conclusive and cannot be taken to
rule out the notion of gravitational energy-momentum density.

\section{A regularized expression for the gravitational 
energy-momentum}

Let us briefly recall the Lagrangian formulation of the TEGR.
The Lagrangian density for the gravitational field in the TEGR in
empty space-time is given by

\begin{eqnarray}
L(e_{a\mu})&=& -k\,e\,({1\over 4}T^{abc}T_{abc}+
{1\over 2} T^{abc}T_{bac} -T^aT_a) \nonumber \\
&\equiv&-k\,e \Sigma^{abc}T_{abc}\;,
\label{3}
\end{eqnarray}
where $k=1/(16 \pi)$ and $e=\det(e^a\,_\mu)$. The tensor
$\Sigma^{abc}$ is defined by

\begin{equation}
\Sigma^{abc}={1\over 4} (T^{abc}+T^{bac}-T^{cab})
+{1\over 2}( \eta^{ac}T^b-\eta^{ab}T^c)\;,
\label{4}
\end{equation}
and $T^a=T^b\,_b\,^a$. The quadratic combination
$\Sigma^{abc}T_{abc}$ is proportional to the scalar curvature
$R(e)$, except for a total divergence. The field
equations for the tetrad field read

\begin{equation}
e_{a\lambda}e_{b\mu}\partial_\nu(e\Sigma^{b\lambda \nu})-
e(\Sigma^{b \nu}\,_aT_{b\nu \mu}-
{1\over 4}e_{a\mu}T_{bcd}\Sigma^{bcd})
\;=0 \;,
\label{5}
\end{equation}
It is possible to prove by explicit calculations that the left hand
side of Eq. (5) is exactly given by ${1\over 2}\,e\,
\lbrack R_{a\mu}(e)-{1\over 2}e_{a\mu}R(e)\rbrack$.
As usual, tetrad fields convert space-time into Lorentz indices and
vice-versa.

The definition for the gravitational energy-momentum has first been
obtained in the Hamiltonian formulation of the TEGR 
\cite{Maluf6,Maluf7}.
However either the Hamiltonian or Lagrangian field equations may be
suitably interpreted as equations that define the gravitational 
energy-momentum. The momentum canonically conjugated to the tetrad
components $e_{aj}$ is given  by $\Pi^{aj}=-4ke\Sigma^{a0j}$. The
latter quantity yields the definition of the 
gravitational energy-momentum $P^a$ contained within a
volume $V$ of the three-dimensional spacelike hypersurface 
\cite{Maluf2,Maluf3,Maluf4},

\begin{equation}
P^a=-\int_V d^3x\, \partial_k \Pi^{ak}\;.
\label{6}
\end{equation}
$P^a$ transforms as a vector under the global SO(3,1) 
group. It describes the gravitational energy-momentum with respect to
observers adapted to $e^a\,_\mu$. These observers are characterized by
the velocity field $u^\mu=e_{(0)}\,^\mu$, and by the acceleration 
$a^\mu$ given by Eq. (2). 

Let us assume that the space-time is asymptotically flat.
The total gravitational energy-momentum is given by

\begin{equation}
P^a=-\oint_{S\rightarrow \infty}  dS_k \,\Pi^{ak}\;.
\label{7}
\end{equation}
The field quantities are evaluated on a surface $S$ in the limit
$r \rightarrow \infty$. 

In Eqs. (6,7) it is implicitly assumed that the reference
space is determined by a set of tetrad fields $e^a\,_\mu$ for the
flat space-time such that the condition $T^a\,_{\mu\nu}=0$ is 
satisfied. However in general there exist flat space-time tetrad 
fields for which $T^a\,_{\mu\nu}\ne 0$. In this case we may generalize
Eq. (6) by adding a suitable reference space subtraction term, exactly
like in the Brown-York formalism \cite{BY}. The Brown-York quasi-local
energy expression is regularized by subtracting the energy of a flat
slice of the flat space-time.

Let us denote 
$T^a\,_{\mu\nu}(E)=\partial_\mu E^a\,_\nu-\partial_\nu E^a\,_\mu$,
and $\Pi^{aj}(E)$ as the expression of $\Pi^{aj}$ constructed out of 
{\it flat tetrads} $E^a\,_\mu$. The regularized form of the 
gravitational energy-momentum $P^a$ is defined by

\begin{equation}
P^a=-\int_V d^3x\,\partial_k\lbrack\Pi^{ak}(e) - \Pi^{ak}(E)\rbrack\;.
\label{8}
\end{equation}
This definition guarantees that the energy-momentum of the flat 
space-time always vanishes. The reference space-time is determined by 
the tetrad fields $E^a\,_\mu$, obtained from $e^a\,_\mu$ by requiring
the vanishing of the physical parameters like mass, angular momentum, 
etc.

The total gravitational energy-momentum is obtained by integrating over
the whole three-dimensional spacelike section. Assuming again that the 
space-time is asymptotically flat, we have

\begin{equation}
P^a=-\oint_{S\rightarrow \infty}  dS_k
\,\lbrack\Pi^{ak}(e) - \Pi^{ak}(E)\rbrack\;,
\label{9}
\end{equation}
where the surface $S$ is established at spacelike infinity.
Like Eq. (6), the definition above transforms as a vector under the
global SO(3,1) group. 

The definition given by Eq. (8) is valid also in the context of 
space-times with an arbitrary topology. It is legitimate to take
the tetrad fields $E^a\,_\mu$ to represent the pure de Sitter or
anti-de Sitter spaces, for instance, in which case Eq. (8)
represents the gravitational energy-momentum defined about the
latter space-times.

\section{Reference frames in the Kerr space-time and the total
gravitational energy}

In this section we will apply Eq. (9) to a simple set of tetrad
fields that describes the Kerr space-time, in order to illustrate
the procedure (of course the analysis of the gravitational energy of 
the Kerr space-time may be carried out by means of several 
approaches). For this purpose  we will evaluate the total 
gravitational energy. The asymptotic form of the Kerr metric tensor
describes the exterior region of a rotating isolated material
system. The set of tetrad fields to be considered
allows a straightforward evaluation of connections and curvature 
(in the context of Riemannian geometry), but it has neither a simple 
geometrical structure when written in cartesian coordinates, nor an
appropriate asymptotic behaviour. Before 
we carry out this analysis, we will recall the construction of 
tetrad fields as reference frames. We start by considering the flat
Minkowski space-time with cartesian coordinates $x^\mu$.

Besides $x^\mu$, the flat space-time is endowed with 
cartesian coordinates $q^a$. The coordinate system $q^a$ establishes
a global reference frame. The transformation matrix that relates
the two coordinate systems defines a set of tetrad fields for the
Minkowski space-time, $E^a\,_\mu=\partial_\mu q^a$. The coordinate
transformation $dq^a= E^a\,_\mu dx^\mu$ can be globally integrated,
and therefore it establishes a holonomic transformation between $q^a$
and $x^\mu$. Rotations and boosts between $q^a$ and $x^\mu$ are the 
two basic SO(3,1) transformations. The condition 

\begin{equation}
E_{(i)j}(t,x,y,z)=E_{(j)i}(t,x,y,z)\,,
\label{10}
\end{equation}
ensures that $q^a$ is not rotating with respect to $x^\mu$, because a
rotation will give rise to antisymmetric components in the sector 
$E_{(i)j}$ (Latin indices from the middle of the alphabet run
from 1 to 3). On the other hand, a boost between $q^a$ and $x^\mu$ 
implies that $E^{(0)}\,_k\ne 0$ \cite{Maluf2}. Therefore by imposing 
$E^{(0)}\,_k=0$, or, equivalently, the time gauge condition,

\begin{equation}
E_{(i)}\,^0=0\,,
\label{11}
\end{equation}
we ensure that the two coordinate systems have a unique time scale. 
Both $q^a$ and $x^\mu$ describe the flat space-time. Thus
we may say that if conditions (10) and (11) are imposed on 
$E^a\,_\mu(t,x,y,z)$ the reference space-time with coordinates 
$q^a$ is neither rotating nor undergoing a boost with respect to the 
space-time with coordinates $x^\mu$ \cite{Maluf2}. If these 
conditions are imposed we have $E^a\,_\mu(t,x,y,z)$ 
$=\delta^a_\mu$,
or

\begin{equation}
E^a\,_\mu(t,r,\theta,\phi)=
\pmatrix{1&0&0&0\cr
0&\sin\theta\, \cos\phi & r\,\cos\theta\,\cos\phi
& -r\,\sin\theta\,\sin\phi\cr
0&\sin\theta\, \sin\phi & r\,\cos\theta\,\sin\phi
&  r\,\sin\theta\,\cos\phi\cr
0&\cos\theta & -r\,\sin\theta & 0\cr}\;.
\label{12}
\end{equation}

From a different but equivalent point of view we may say that\par
\noindent $E^a\,_\mu(t,x,y,z)=\delta^a_\mu$ 
is adapted to stationary
observers in space-time, namely, observers that are endowed with 
the velocity field $u^\mu= E_{(0)}\,^\mu=\delta^\mu_{(0)}$ and 
acceleration $a^\mu$ given by Eq. (2). In this case, $a^\mu=0$. 

A geometrical interpretation of tetrad fields as an observer's 
frame can be given as follows. We consider an arbitrary path
$x^\mu(s)$ of the observer 
in Minkowski space-time, where $s$ is the 
proper time of the observer.
We identify $dx^\mu/ds=u^\mu= E_{(0)}\,^\mu$, where 
$E_{(0)}\,^\mu$ is the timelike component of the orthonormal 
frame (the temporal axis of the observer's local frame).
According to the {\it hypothesis of locality} \cite{Mashh1},
a noninertial observer at each instant along its worldline is 
equivalent to an otherwise identical momentarily comoving inertial 
observer. It follows from the hypothesis of locality that each 
noninertial observer is endowed with an orthonormal tetrad frame
$E_a\,^\mu$, whose derivative along the path is
given by \cite{Mashh2,Mashh3}

\begin{equation}
{{dE_a\,^\mu} \over {ds}}=\phi_a\,^b\,E_b\,^\mu\,,
\label{13}
\end{equation}
where $\phi_{ab}$ is the antisymmetric acceleration tensor (not to
be confused with $\phi^{aj}$ given by Eq. (28)). According to Refs.
\cite{Mashh2,Mashh3},
in analogy with the Faraday tensor we can identify
$\phi_{ab} \rightarrow (-{\bf a}, {\bf \Omega})$, where ${\bf a}$ is
the translational acceleration ($\phi_{(0)(i)}=a_{(i)}$)
and ${\bf \Omega}$ is the frequency of
rotation of the local spatial frame  with respect to a nonrotating
(Fermi-Walker transported \cite{Hehl}) frame. 
The invariants constructed out of
$\phi_{ab}$ establish the acceleration scales and lengths 
\cite{Mashh1}. It follows from Eq. (13) that

\begin{equation}
\phi_a\,^b= E^b\,_\mu {{dE_a\,^\mu} \over {ds}}=
E^b\,_\mu \,u^\lambda\nabla_\lambda E_a\,^\mu\,.
\label{14}
\end{equation}

Therefore given any set of tetrad fields for an arbitrary 
gravitational field configuration its geometrical interpretation 
can be obtained by suitably interpreting the velocity field 
$u^\mu=\,e_{(0)}\,^\mu$ and the acceleration tensor $\phi_{ab}$,
in case we ``switch off" the gravitational field by making
$e^a\,_\mu \rightarrow E^a\,_\mu$. 
In several situations it turns out to be
easy to impose conditions (10) and (11) on $e^a\,_\mu$. However, 
the proper interpretation of $\phi_{ab}$ along a typical 
trajectory determined by the velocity vector $u^\mu$ of a class 
of observers adapted to a tetrad field seems to be a condition 
stronger than Eqs. (10) and (11).

Now we consider the Kerr space-time. The line element is given by

\begin{equation}
ds^2=-{\psi^2 \over\rho^2}dt^2-
{{2\chi \sin^2\theta}\over \rho^2}d\phi\,dt
+{\rho^2 \over \Delta}dr^2+
\rho^2 d\theta^2+
{{\Sigma^2 \sin^2\theta}\over \rho^2}d\phi^2 \;,
\label{15}
\end{equation}
where 

$$\rho^2=r^2+a^2 cos^2\theta\,,$$ 

$$\Delta= r^2 +a^2 -2mr\,,$$

$$\chi=2amr\,,$$

$$\Sigma^2=(r^2+a^2)^2-\Delta a^2\,\sin^2\theta\,,$$

$$\psi^2=\Delta -a^2\,\sin^2\theta\,.$$

\noindent 
Imposition of conditions (10) and (11) yields the following 
expression for $e^a\,_\mu$,

\begin{equation}
e_{a\mu}=\pmatrix{
-{1\over \rho}\sqrt{\psi^2+{\chi^2\over \Sigma^2}\sin^2\theta} &
0&0&0\cr
{\chi \over {\Sigma \rho}}\sin\theta\,\sin\phi &
{\rho \over \sqrt{\Delta}}\sin\theta\,\cos\phi &
\rho\,\cos\theta\,\cos\phi &
-{\Sigma \over \rho} \sin\theta\,\sin\phi\cr
-{\chi \over{\Sigma \rho}}\sin\theta\,\cos\phi &
{\rho \over \sqrt{\Delta}}\sin\theta\,\sin\phi &
\rho\,\cos\theta\,\sin\phi &
{\Sigma \over \rho}\sin\theta\,\cos\phi\cr
0&{\rho \over \sqrt{\Delta}}\cos\theta&-\rho\,\sin\theta&0\cr}
\,.
\label{16}
\end{equation}
The transformation $dq^a=e^a\,_\mu dx^\mu$ determined by the expression
above cannot be globally integrated, because in this case
$e^a\,_\mu \ne \partial_\mu q^a$. Therefore $dq^a=e^a\,_\mu dx^\mu$ is
an anholonomic transformation. An important feature of the equation
above is that its expression in the asymptotic limit 
$r \rightarrow \infty$ is given by Eq. (1). Thus we may say that Eq. 
(16) is adapted to stationary observers at spacelike infinity. We also
note that the flat space-time limit of Eq. (16) yields Eq. (12), and
therefore $T^a\,_{\mu\nu}(E)=0$.

Equation (16) above has proven to describe satisfactorily the
energy-momentum properties of the Kerr space-time \cite{Maluf2}
(we note that tetrad fields for the Kerr space-time have also been
addressed in Refs. \cite{Baekler,Pereira}).
However, the line element given by Eq. (15) admits a simple form
that is useful for computational purposes, and which reads 

\begin{equation}
e_{a\mu}=\pmatrix{-A&0&0&0\cr
0&{\rho \over \sqrt{\Delta}}&0&0\cr
0&0&\rho&0\cr
B&0&0&C \cr}\,,
\label{17}
\end{equation}
where

\begin{eqnarray}
A&=&\biggl( 
{{\chi^2\sin^2\theta+\psi^2 \Sigma^2}\over {\rho^2 \Sigma^2}}
\biggr)^{1\over 2}\,, \nonumber \\
B&=&-{{\chi \sin\theta}\over {\rho\Sigma}}\,, \nonumber \\
C&=&{{\Sigma \sin\theta}\over{\rho}}\,.
\label{18}
\end{eqnarray}

The flat space-time limit of Eq. (17) is given by

\begin{equation}
E^a\,_\mu=\pmatrix{1&0&0&0\cr
0&1&0&0\cr
0&0&r&0\cr
0&0&0&r\sin\theta\cr}\,.
\label{19}
\end{equation}
The expression above yields three nonvanishing torsion 
components:\par
\noindent $T_{(2)12}(E)=1$, $T_{(3)13}(E)=\sin\theta$, and 
$T_{(3)23}(E)=r\cos\theta$. 
Inspite of its simplicity, this tetrad field has a rather intricate
structure when written in cartesian coordinates. It reads

\begin{equation}
E^a\,_\mu(t,x,y,z)=\pmatrix{1&0&0&0\cr
0&
{x\over r}&
{y\over r}&
{z\over r}\cr
0&
{{xz}\over {r\sqrt{x^2+y^2}}}&
{{yz}\over {r\sqrt{x^2+y^2}}}&
-{{\sqrt{x^2+y^2}}\over r}\cr
0&
-{y \over {\sqrt{x^2+y^2}}}&
{x\over {\sqrt{x^2+y^2}}}&
0\cr}\,.
\label{20}
\end{equation}
In view of the geometrical structure of the equation above, we see
that, differently from Eq. (16), Eq. (17) does not display the 
asymptotic behaviour determined by Eq. (1). Moreover, in general
the tetrad
field determined by Eq. (20) is adapted to accelerated observers.
In order to verify this fact, let us consider a boost in the $x$
direction, say, of Eq. (20). We find

\begin{equation}
E^a\,_\mu(t,x,y,z)=\pmatrix{\gamma& -\beta\gamma&0&0\cr
-\beta\gamma&
\gamma{x\over r}&
{y\over r}&
{z\over r}\cr
0&
{{xz}\over {r\sqrt{x^2+y^2}}}&
{{yz}\over {r\sqrt{x^2+y^2}}}&
-{{\sqrt{x^2+y^2}}\over r}\cr
0&
-{y \over {\sqrt{x^2+y^2}}}&
{x\over {\sqrt{x^2+y^2}}}&
0\cr}\,,
\label{21}
\end{equation}
where $\beta$ and $\gamma$ are constants defined by $\beta=v/c$ and
$\gamma=\sqrt{1-\beta^2}$. It is easy to see that along 
an observer's trajectory whose velocity is 
determined by $u^\mu=(\gamma,-\beta\gamma,0,0)$ the quantities 
$\phi_{(j)}\,^{(k)}=u^i(E^{(k)}\,_m \partial_i E_{(j)}\,^m)$ 
constructed out of Eq. (21) are
nonvanishing. This fact indicates that along the observer's path the
spatial axis $E_{(i)}\,^\mu$ rotate. Nevertheless Eq. (17)
yields a satisfactory value for the total gravitational 
energy-momentum, as we will see.

We will integrate Eq. (9) over a surface of constant radius 
$x^1=r$, and then we require $r\rightarrow \infty$. Therefore 
we make $k=1$ in Eq. (9). Out of Eq. (17) we evaluate all torsion 
components $T_{a\mu\nu}$. We need the quantity 

$$\Sigma^{(0)01}=e^{(0)}\,_0\Sigma^{001}={1\over 2}
e^{(0)}\,_0(T^{001}-g^{00}T^1)\,.$$
The calculations are lengthy but straightforward, and therefore they
will be omitted here. We find

\begin{equation}
-\Pi^{(0)1}(e)=4ke\Sigma^{(0)01}=
-{1\over {8\pi}}{\sqrt{\Delta}\over\rho}
(\partial_r\Sigma)\sin\theta\,.
\label{22}
\end{equation}
The expression of $\Pi^{(0)1}(E)$ is obtained from Eq. (19) or,
equivalently, by just making
$m=a=0$ in the expression above. It is given by

\begin{equation}
\Pi^{(0)1}(E)={1\over {4\pi}}r\sin\theta\,.
\label{23}
\end{equation}
Thus the gravitational energy contained within a surface $S$
of constant radius $r$ reads

\begin{eqnarray}
P^{(0)}&=&-\oint_S  dS_k
\,\lbrack \Pi^{(0)k}(e) - \Pi^{(0)k}(E)\rbrack \nonumber \\
&=&\int_S d\theta d\phi\, {1\over {4\pi}} \sin\theta\biggl(
-{1\over 2}{\sqrt{\Delta}\over\rho}
(\partial_r\Sigma) + r\biggr)\,.
\label{24}
\end{eqnarray}

In the limit $r\rightarrow \infty$ we have

\begin{equation}
4ke\Sigma^{(0)01}\cong -{1\over {4\pi}}r\sin\theta(1-{m\over r})\,.
\label{25}
\end{equation}
Therefore for the total gravitational energy of the Kerr space-time
we obtain

\begin{equation}
P^{(0)}\cong
\int_{r\rightarrow \infty}d\theta d\phi\,{1\over{4\pi}}\sin\theta
\biggl(-r(1-{m\over r})+r\biggr) =m\,,
\label{26}
\end{equation}
which is the expected result.

We may also integrate Eq. (24) on the surface of constant radius 
$r=r_+$, where $r_+$ is the external horizon of the Kerr black hole. 
On this surface the function $\Delta= r^2 +a^2 -2mr$ vanishes. 
Therefore we find $P^{(0)}=r_+=m+\sqrt{m^2-a^2}$, a result that is 
quite different from the irreducible mass of que Kerr black hole.
The localization of gravitational energy in the Kerr space-time is 
correctly described by Eq. (16), according to the discussion
in Ref. \cite{Maluf2}. 
As discussed above, the frame determined by Eq. (16) 
is adapted to stationary observers at spacelike infinity.

Before we close this section let us recall that
by means of simple algebraic manipulations 
an expression for the gravitational energy-momentum flux
was developed in Ref. \cite{Maluf3}. 
This expression follows directly from the
field equations (5). It reads

\begin{equation}
{d \over {dt}}\biggl[ -\int_V d^3x\,\partial_j \Pi^{aj}\biggr]=
-\oint_S dS_j \, \phi^{aj}\;,
\label{27}
\end{equation}
where

\begin{equation}
\phi^{aj}=k [ ee^{a\mu}(4\Sigma^{bcj}T_{bc\mu}-
\delta^j_\mu \Sigma^{bcd}T_{bcd}) ]\;.
\label{28}
\end{equation}
The quantity above represents the $a$ component of the flux density
in the $j$ direction. In Ref. \cite{Maluf4} this formalism was
applied to the evaluation of energy loss in Bondi's radiative
space-time. In Eqs. (27) and (28) it is assumed that for the flat
space-time we have $T_{a\mu\nu}(E)=0$. We may address
Eq. (27) in the context of the present analysis. Let us
assume that $T_{a\mu\nu}(E) \ne0$.
Since $E^a\,_\mu$ is also a solution of the
field equations (5), Eq. (27) is trivially satisfied for
$E^a\,_\mu$. Therefore we may write

\begin{equation}
{d \over {dt}}\biggl[ -\int_V d^3x\,
\partial_j\lbrack \Pi^{aj}(e) - \Pi^{aj}(E)\rbrack \biggr]=
-\oint_S dS_j \lbrack \phi^{aj}(e)- \phi^{aj}(E)\rbrack   \;,
\label{29}
\end{equation}
where $\phi^{aj}(E)$ is constructed out of $E^a\,_\mu$. We
observe that as long as $E^a\,_\mu$ (and consequently
$\Pi^{aj}(E)$) is time independent, the left hand side of Eq. (29)
is simplified and therefore the 
energy-momentum loss can be easily calculated out of any
set of tetrad fields. The vanishing of $\phi^{aj}(e)-
\phi^{aj}(E)$ at spacelike infinity (a feature that is 
expected to take place for asymptotically flat space-times) 
ensures the conservation of the total gravitational 
energy-momentum.

\section{Discussion}

In this article we have extended the definition for the 
gravitational energy-momentum previously considered in the framework 
of the TEGR, which requires $T_{a\mu\nu}(E)=0$ for the flat 
space-time, to the case where the flat space-time tetrad fields 
$E^a\,_\mu$ yield $T_{a\mu\nu}(E)\ne 0$. In the context of the 
regularized gravitational energy-momentum definition it is not 
strictly necessary to stipulate asymptotic boundary conditions for 
tetrad fields that describe asymptotically flat space-times.

We have seen that Eqs. (13) and (14) provide a physical
interpretation for a set of tetrad fields in Minkowski space-time,
in terms of the linear acceleration and rotation of an observer 
adapted to the frame $E^a\,_\mu$, endowed with velocity 
$u^\mu=E_{(0)}\,^\mu$. We note that all frames obtained from
$E^a\,_\mu=\delta^a_\mu$ by means of a global SO(3,1) transformation
(determined by constant transformation matrices $\Lambda^a\,_b$) 
yield $\phi_a\,^b=0$, according to Eq. (14). Thus the requirement 
$\phi_{ab}=-\phi_{ba}=0$ seems to be  equivalent to conditions 
(10) and (11).

The definition given by Eq. (8) can be applied to an arbitrary 
volume $V$ in any space-time, with an arbitrary topology. We 
propose that Eq. (8) represents the gravitational energy-momentum
relative to the frame determined by the tetrad field $e^a\,_\mu$,
with $E^a\,_\mu$ representing the tetrad field when the 
physical parameters of the metric tensor (mass, angular momentum, 
etc.) vanish. 

\bigskip
\bigskip
\noindent {\bf Acknowledgements}\par
\noindent This work was partially supported by the Brazilian Agency 
CNPQ. M. V. O. Veiga is supported by a CNPQ fellowship.\par

\end{document}